# Experimental Validation of Sensor Fusion-based GNSS Spoofing Attack Detection Framework for Autonomous Vehicles

**Sagar Dasgupta***
Department of Civil, Construction, and Environmental Engineering
The University of Alabama, Tuscaloosa, AL 35487
Email: sdasgupta@crimson.ua.edu

**Kazi Hassan Shakib**
Department of Civil, Construction, and Environmental Engineering
The University of Alabama, Tuscaloosa, AL 35487
Email: khshakib@crimson.ua.edu

**Mizanur Rahman, Ph.D.**
Department of Civil, Construction, and Environmental Engineering
The University of Alabama, Tuscaloosa, AL 35487
Email: mizan.rahman@ua.edu

## ABSTRACT

In this paper, we validate the performance of the a sensor fusion-based Global Navigation Satellite System (GNSS) spoofing attack detection framework for Autonomous Vehicles (AVs). To collect data, a vehicle equipped with a GNSS receiver, along with Inertial Measurement Unit (IMU) is used. The detection framework incorporates two strategies: The first strategy involves comparing the predicted location shift, which is the distance traveled between two consecutive timestamps, with the inertial sensor-based location shift. For this purpose, data from low-cost in-vehicle inertial sensors such as the accelerometer and gyroscope sensor are fused and fed into a long short-term memory (LSTM) neural network. The second strategy employs a Random-Forest supervised machine learning model to detect and classify turns, distinguishing between left and right turns using the output from the steering angle sensor. In experiments, two types of spoofing attack models: turn-by-turn and wrong turn are simulated. These spoofing attacks are modeled as SQL injection attacks, where, upon successful implementation, the navigation system perceives injected spoofed location information as legitimate while being unable to detect legitimate GNSS signals. Importantly, the IMU data remains uncompromised throughout the spoofing attack. To test the effectiveness of the detection framework, experiments are conducted in Tuscaloosa, AL, mimicking urban road structures. The results demonstrate the framework's ability to detect various sophisticated GNSS spoofing attacks, even including slow position drifting attacks. Overall, the experimental results showcase the robustness and efficacy of the sensor fusion-based spoofing attack detection approach in safeguarding AVs against GNSS spoofing threats.

## BACKGROUND AND MOTIVATION

Autonomous Vehicles (AVs) heavily depend on the Global Navigation Satellite System (GNSS) for accurate and continuous real-time position information, crucial for their autonomous navigation, security, and safety-critical applications. GNSS comprises multiple systems owned by different countries, including GPS (USA) (*2*), BeiDou/BDS (China) (*3*), Galileo (Europe) (*4*), GLONASS (Russia) (*5*), IRNSS/NavIC (India) (*6*), and QZSS (Japan) (*7*). Despite their importance, GNSS signals are relatively weak, often likened to the visibility of a 25-watt light bulb from a distance of about 20,000 kilometers, making them susceptible to interference. While GPS provides both encrypted and unencrypted signals, the encrypted signals are accessible only to government and military users, leaving unencrypted civilian signals vulnerable to both intentional and unintentional threats. To meet the navigation requirements of AVs, precise location information is essential, with positioning errors needing to be within the centimeter level. Currently, AVs employ a differential global positioning system (DGPS) for correcting localization solutions. DGPS can achieve position error as low as 0.01m (*8*) by utilizing a reference station alongside regular GPS signal. Throughout the rest of this paper, DGPS will be referred to as both GNSS and GPS.

The performance of the GNSS, is affected by physical degradation of the radio signal caused by natural, unintentional, and intentional threats (*9*). As GNSS signals travel through the Earth's atmosphere to reach an AV's GNSS receiver, their strength diminishes. Additionally, tall structures reflecting these signals introduce positional errors and disrupt continuous signal availability at the receiver (*10–14*), creating unintentional threats. On the other hand, intentional threats involve attackers blocking the GNSS signal or transmitting fake signals. International threats can be broadly categorized into two types: jamming and spoofing. In jamming (*15*), a powerful GNSS signal is transmitted to prevent the genuine signal from reaching the target GNSS receiver. However, spoofing (*16–20*) is a more sophisticated attack where the attacker manipulates the authentic GNSS signal structure, transmitting false location information to the target AV. The AV, unaware of the manipulation, updates its navigation route based on the spoofed signal (*21*). Such sophisticated GNSS spoofing attacks require the spoofer to obtain the target vehicle's current location, destination, and route information. In this experiment, the sensor fusion-based GNSS spoofing attack detection framework is tested against sophisticated spoofing attacks.

One of the main objectives of manipulating a GNSS receiver during a spoofing attack is to interfere with the GNSS signal, potentially redirecting a target AV to the wrong destination, posing risks to the safety of passengers and transportation of goods. Spoofing attacks are categorized into three types: simplistic, intermediate, and sophisticated (*22*). Simplistic attacks involve using a commercial GPS signal simulator along with a power amplifier and antenna (*23*). These asynchronous attacks are akin to signal jamming, causing the GPS receiver to lose lock and go through partial or complete reacquisition. Detecting such attacks is relatively straightforward. Intermediate attacks utilize portable receiver spoofers, made feasible by the advent of software-defined radio (SDR) technology (*24*). This development has simplified the creation of these portable spoofers, enabling intermediate-level attacks. In this scenario, the attacker closely tracks the victim AV to gather information about its GPS receiver antenna's position and velocity, ensuring precise positioning of the spoofed signals relative to the legitimate signals at the AV's antenna. As these attacks are synchronous, they pose greater challenges for detection. Sophisticated attacks represent the highest level of sophistication, employing multiple phase-locked portable receiver-spoofers. These attacks are specifically designed to deceive angle-of-arrival-based defensive systems, making them highly formidable and difficult to counter. Unfortunately, due to the dynamic nature of spoofing attacks, no single detection method can identify all types of GNSS spoofing attacks (*25*). Therefore, researchers focus on developing methods that increase the difficulty of executing a spoofing attack, aiming to enhance the resilience of GNSS systems against these threats. The detection framework operates at the localization solution level, making it sufficient to model the attack by simply replacing the legitimate localization data (latitude, longitude, and altitude) with spoofed locations without actually spoofing the GPS receiver antenna.

In this paper, we validate the performance of our previously proposed sensor fusion-based Global Navigation Satellite System (GNSS) spoofing attack detection framework designed specifically for Autonomous Vehicles (AV) (*1*). The detection framework employs two strategies: First, it compares the predicted location shift (distance between consecutive timestamps) with the inertial sensor-based location shift, monitoring vehicle motion states. Data from low-cost in-vehicle inertial sensors are fused into an LSTM neural network to predict AV travel distance. Second, a Random-Forest model detects and classifies turns based on the steering angle sensor. Both strategies compare GNSS-derived speed with vehicle velocity. Two types of spoofing attacks (turn-by-turn, wrong turn) are simulated as SQL injections, with the IMU data remaining uncompromised. Experiments in Tuscaloosa, AL, mimic urban roads to test the framework's effectiveness.

The paper follows the following structure: The "Attack Model" section introduces the attack scenarios and the process of creating the attacks. Next, the "Attack Detection Framework" section describes the attack models and the adapted attack detection framework utilized in the experiments. The "Experimental Setup" section provides an overview of the experimental arrangement. In the "Detection Framework Performance" section, the performance of the detection framework is analyzed. Lastly, the "Conclusion" section offers concluding remarks and highlights potential directions for future research.

## ATTACK MODEL

Throughout the experiment, we simulate two types of GNSS spoofing scenarios (turn-by-turn, wrong turn)(See **Figure 1**). The spoofing detection model introduced in this paper is solely dependent on the latitude and longitude data from the GNSS, without considering any other signal parameters. Therefore, by substituting the legitimate latitude and longitude data in the MongoDB database with spoofed latitude and longitude values (See **Figure 2**), we can accurately mimic a spoofing attack and effectively test the detection framework's performance against such attacks. In these experiments, a spoofed database is created. The spoofed database contains the latitude and longitude of spoofed routes for both spoofing scenarios. The spoofed route is selected based on the assumption that an attacker possesses information about the probable route of the victim AV before initiating the attack. This knowledge allows the attacker to gradually manipulate the perceived location of the AV, making it challenging to detect spoofing. To create and obtain latitude and longitude data along the route, ArcGIS's Network Analyst tool is utilized. To replicate the movement of an actual vehicle, the update rate of the spoofed route's location is determined based on the average speed observed on that specific roadway. For the testing routes, it is assumed that the average vehicle speed is 25 mph, equivalent to 11.18 m/s. Considering the AV's GNSS data output frequency to be 1 Hz, latitude and longitude points along the spoofed route are generated approximately 11 meters apart from each other. Once the attacker successfully spoofs the AV's GNSS receiver, the latitude and longitude data columns in the database are substituted with the spoofed route data from the spoofed database. Consequently, the AV will perceive the spoofed location as authentic. However, it is important to note that during the attack, all other sensor data remains uncompromised. This means that all other data columns are continuously updated in real-time with legitimate data from the CPT7700, maintaining their integrity throughout the spoofing attack.

A set of spoofed routes are generated for each spoofing attack scenario. For the turn-by-turn attack data, a route is created wherein the AV's location is shifted from its current position. The shift is directed towards the adjacent parallel road, with consideration for the high density of the surroundings. As a result, the shift is intentionally kept relatively small. To create wrong turn attacks seven intersections are chosen, and four wrong right turn and three wrong left turn routes are created. Whenever the AV is near one of the chosen intersections, a spoofed route is injected into the database to simulate the attack. For all the spoofing attacks, the injection of the spoofed database is done by SQL injection attack (*26, 27*) technique that allows the attacker to add, delete, and modify the database contents.

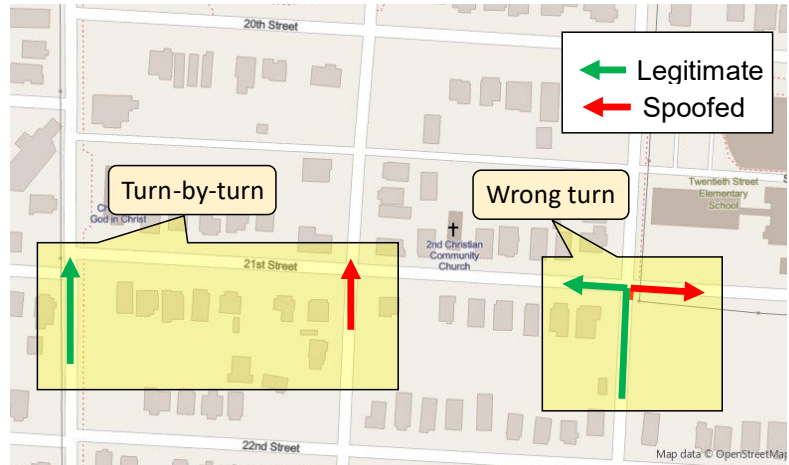

**Figure 1 Modeled spoofing attacks (samples)**

## ATTACK DETECTION FRAMEWORK

The GNSS spoofing attack detection framework (*1*) utilizes two concurrent strategies, fusing data from in-vehicle sensors—GNSS, accelerometer, gyroscope, and speedometer to achieve a unified and resilient GNSS spoofing attack detection approach. The first strategy focuses on developing a vehicle state prediction model by training a deep recurrent neural network (LSTM) with attack-free speed, acceleration, and gyroscope data to predict the location shift between consecutive timestamps. This strategy also continuously monitors the vehicle's motion state using the speedometer output. In the second strategy, gyroscope data is employed to recognize left and right turns. A Random Forest (RF) algorithm is trained with vehicle route data to learn patterns of left and right turns, enabling the detection and classification of turning maneuvers. The turn detection strategy considers both inertial sensor output and GNSS turning information and uses input from the speedometer sensor to distinguish such cases.

The experimental framework is presented in **Figure 2**. A training dataset is created to test all the detection models using the setup explained in the previous paragraph. The driving route of the training dataset replicates an urban road structure, aiming to imitate real-world scenarios. The GNSS traces (latitude, and longitude) is presented in **Figure 3**. Data from the CPT7700's IMU, including X, -Y, Z acceleration, and X, -Y, Z gyro data, along with speed data (north, east, and up velocity) and location data (latitude and longitude), are stored in a MongoDB database. MongoDB is chosen for its superior performance, scalability, availability, and flexibility compared to SQL databases. (*28*). The frequency of data is 1Hz. The Haversine formula is utilized to determine the distance traveled between two consecutive timestamps based on the data obtained from GNSS. (See **Equation 1**) (*29*):

$$d = 2r \, \sin^{-1}(\sqrt{\sin^2\left(\frac{\varphi_2 - \varphi_1}{2}\right) + \cos(\varphi_1)\cos(\varphi_2)\sin^2(\frac{\psi_2 - \psi_1}{2})}) \qquad (1)$$

where d represent the distance in meters between two points on the Earth's surface; r denotes the Earth's radius (6378 km); $\varphi_1$ and $\varphi_2$ represent the latitudes in radians; and $\psi_1$ and $\psi_2$ denote the longitudes in radians of two consecutive time stamps.

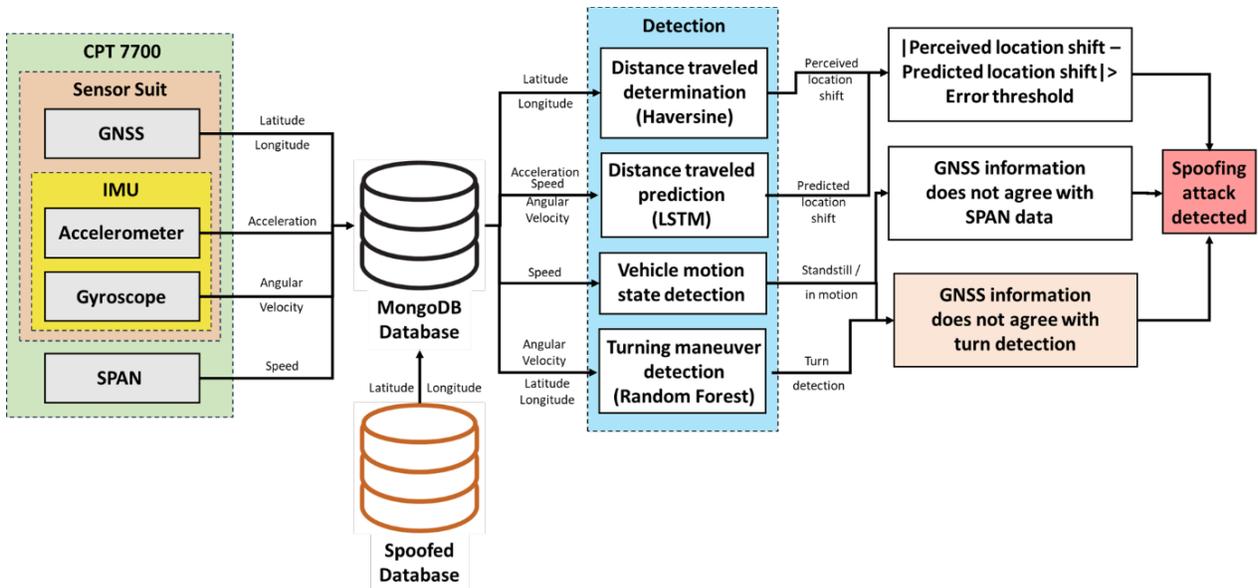

**Figure 2 Experimental detection framework**

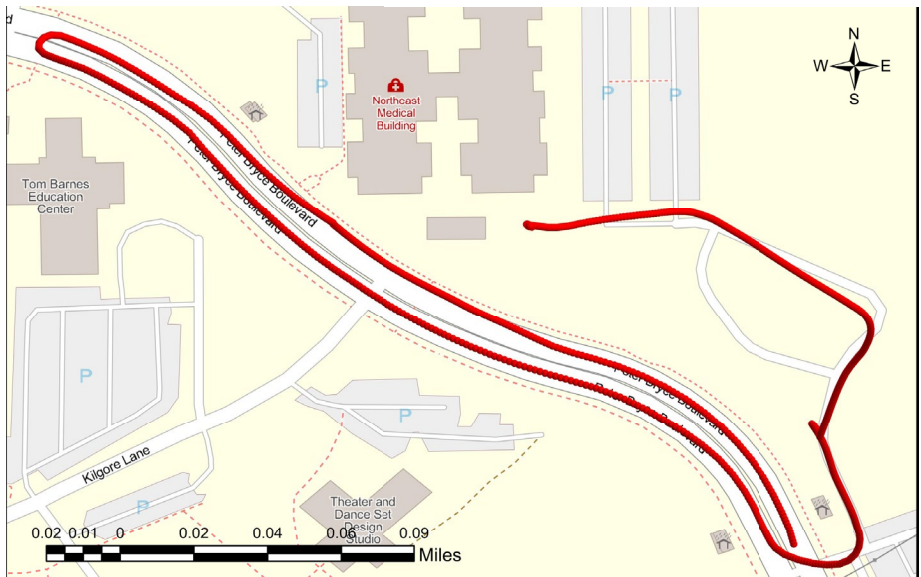

**Figure 3 GNSS traces from the training dataset**

The prediction of the distance traveled by an AV between two consecutive timestamps is achieved using an LSTM architecture with 50 neurons (*30*). The training and validation datasets include X, -Y, Z acceleration, X, -Y, Z gyro, and east, north, and up-speed data. The model's output is the location shift between the current timestamp and the immediate future timestamp. The data generation frequency is 1 Hz, resulting in a time difference of 1 second between two consecutive timestamps. For training purposes, the continuous driving data from the training dataset (see **Figure 3**) is split into training with 7,193 observations and validation with 3,083 observations. Prior to feeding the sensor output into the LSTM training, the input features are normalized between 0 and 1. To select the appropriate LSTM hyperparameters, such as the number of neurons, number of epochs, batch size, and learning rate, a Grid Search approach is employed due to the model's time series-based nature (*30*). The hyperparameter values and the optimizer's name are listed in **Table 1**. After evaluating the LSTM-based prediction model, the Root Mean Square Error (RMSE) of the predicted location shift is measured to be 0.02 m, with the maximum absolute error being 0.06 m.

**TABLE 1 LSTM model hyperparameters and optimizer**

| Hyperparameters and Optimizer | Value |
|---|---|
| Number if neurons | 50 |
| Number of epochs | 50 |
| Batch size | 12 |
| Learning rate | 0.01 |
| Optimizer | Adam |

**Figure 4** displays the Mean Absolute Error (MAE) loss profile, or learning curve, for the trained LSTM model. The y-axis represents the mean absolute error loss values for both the training and validation datasets, while the x-axis indicates the number of epochs. The learning curve illustrates that initially, the loss is comparatively high; however, the training loss steadily decreases and eventually stabilizes, indicating that the LSTM model is not under-fitted. Additionally, as both the training and validation losses reach a stable state quickly, it confirms that the LSTM model is not overfitted. Furthermore, the training and validation losses are consistently low, which is a positive sign of the model's effectiveness. The initial peak in both the training and validation losses indicates that the model was not well-generalized at that stage. However, as the number of epochs increases, the model becomes more stable and better generalized. It is worth noting that the training and testing data used in the analysis represent real-world driving data on urban routes. This ensures that the neural network model demonstrates generalized behavior within an urban network context.

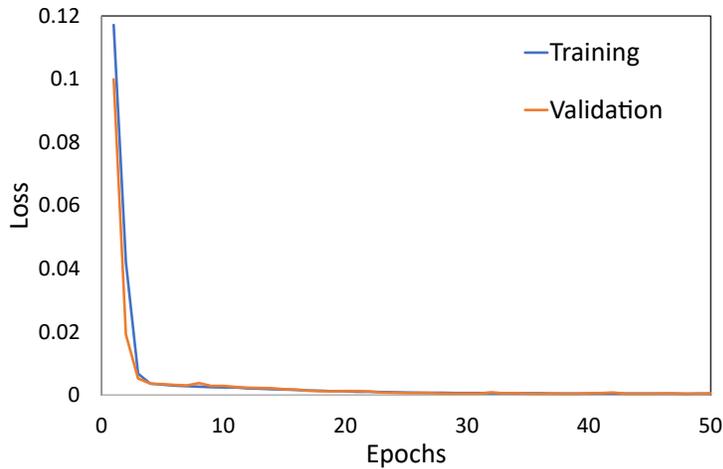

**Figure 4 Comparison of Mean Absolute Error (loss) profiles with the optimal parameter set**

The RF statistical classifier is employed to categorize vehicle maneuvering into three classes: right turn, left turn, and no turn, based on gyroscope output. RF is preferred for its exceptional classification accuracy, capability to model intricate interactions among predictor variables, and its flexibility in conducting various types of statistical data analysis. The RF algorithm constructs classification trees using the dataset and aggregates predictions from all the individual trees to make the final classification decision. X, -Y, and Z gyro data are used as the input features. The training dataset (10026 data points) is labeled as right turn, left turn, and no turn and used as target labels. The training dataset is not balanced in terms of left and right turn data. To solve this problem, Synthetic Minority Oversampling Technique (SMOTE) (*31*) is used to resample the dataset. The SMOTE sampling strategy is set to "auto" so that the minority class is oversampled to achieve an equal number of samples as the majority class. Then the input feature data are normalized between 0 and 1 and fed to the RF model. Cross-validation is performed on the dataset splitting it to 5 folds or subsets. The performance of the trained RF model is presented in **Table 2** in terms of precision, recall, accuracy, and F1 score. Precision is the measure of how accurately a specific turn or no turn is detected out of all observations. The precision of the RF model varies from 93% to 96%. Recall refers to the percentage of the observations where the classification is correct. The recall varies from 89% to 97%. The classification accuracy is 94%. The F1 score reflects the balance between precision and recall. The F1 score ranges from 0.92 to 0.95, which proves that the precision and recall are well-balanced.

**TABLE 2 Random Forest model validation result**

| Turn type | precision | recall | accuracy | f1-score |
|-----------|-----------|--------|----------|----------|
| Right | 0.93 | 0.95 | 0.94 | 0.94 |
| Left | 0.93 | 0.97 | 0.94 | 0.95 |
| No turn | 0.96 | 0.89 | 0.94 | 0.92 |

To compare the gyroscope-based turn detection with the GNSS data. Three consecutive latitude and longitude data are used to determine the turn angle, and based on the turn angle value, the turns are classified. The vehicle's state is determined based on the SPAN data, which consists of north, east, and up velocity components. The RMSE of the SPAN velocity is measured to be 0.03 m/s. If any of the three velocity components (north, east, or up) exceed the threshold of 0.03 m/s, the vehicle state is classified as "moving." Otherwise, if all velocity components are below or equal to 0.03 m/s, the vehicle state is classified as "standstill."

During a trip, all the pre-trained models operate in real-time on the onboard computing unit. The detection algorithms run in parallel. Data from the CPT7700, including GNSS and IMU data, is stored in the MongoDB database. Each model utilizes the required real-time data for spoofing detection. For instance, at timestamp t, the GNSS and IMU data from CPT7700 is immediately fed to the distance traveled determination model to calculate the actual distance traveled between timestamp t and t-1. Simultaneously, the distance traveled prediction model, which is a pre-trained LSTM, predicts the distance traveled between timestamp t and t-1. The results from both models are compared to verify if the GNSS signal is being spoofed. In parallel, the "turning maneuver detection" model leverages the gyroscope data at timestamp t to classify the maneuver type. This information is cross-referenced with the "turn detection" model using data from timestamps t-2, t-1, and t to detect any signs of a spoofing attack in real-time. Furthermore, the vehicle's motion state is determined using the SPAN data at timestamp t. This information is then compared with the GNSS speed output to detect any potential attack occurring in real-time.

**EXPERIMENTAL SETUP**

To test the performance of the sensor fusion-based spoofing attack detection under two attack scenarios (turn-by-turn, wrong turn), a test setup is developed in the Connected and Automated Mobility Laboratory (CAM Lab) of the University of Alabama (See **Figure 5**). A Toyota Camry 2014 SE car is equipped with NovAtel CPT7700 with GNSS and Inertial Navigation System (INS) technology. CPT7700 contains an OEM7700 multi-frequency GNSS receiver, which can track current and upcoming GNSS constellations, including GPS, GLONASS, Galileo, BeiDou, QZSS, and IRNSS. It also includes TerraStar Correction Services with RTK for centimeter-level real-time positioning. It is equipped with NovAtel's SPAN technology for continuous 3D position, velocity, and attitude. The performance of the GNSS receiver is being presented in **Table 3**. CPT7700 is also equipped with the high-performing Honeywell HG4930 Micro Electromechanical System (MEMS) IMU containing gyroscope and accelerometer. The performance of the IMU is being presented in **Table 4**. A CPT7 I/O2 cable is being used to connect CPT7700 with a Windows-based onboard computing unit with the USB port, which supports a hi-speed (480Mb/s) data rate. The Hexagon GNSS-850 tough high-precision antenna with superior tracking performance is being used, featuring multi-point feeding network and radiation pattern optimization technology. It has the ability to track low-elevation satellites while maintaining a high gain for higher-elevation satellites, making it suitable for applications where the sky is partially visible, such as operating close to tree lines, under foliage, or in urban canyons.

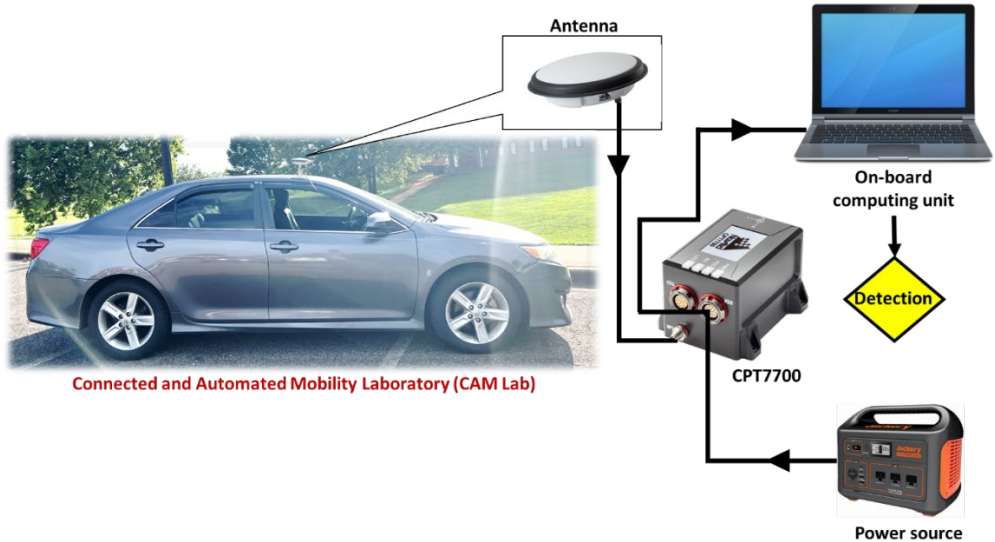

**Figure 5 Hardware setup**

**TABLE 3 GNSS receiver performance**

| Positioning | Accuracy (RMS) |
|---|---|
| Single Point L1 | 1.5 m |
| Single Point L1/L2 | 1.2 m |
| SBAS | 60 cm |
| DGPS (code) | 40 cm |
| TerraStar-C PRO | 2.5 cm |
| TerraStar-L | 40 cm |
| RTK | 1 cm + 1 ppm |

**TABLE 4 IMU performance**

| Gyroscope | | Accelerometer | |
|---|---|---|---|
| Technology | MEMS | Technology | MEMS |
| Dynamic range | 400 °/s | Dynamic range | 20 g |
| Bias instability | 0.45 °/hr | Bias instability | 0.075 mg |
| Angular random walk | 0.06 °/√hr | Velocity random walk | 0.06 m/s/√hr |

## DETECTION FRAMEWORK PERFORMANCE

The attack detection framework is tested against two distinct attack scenarios: turn-by-turn and wrong turn. real-time evaluations are performed, where the attacks are initiated by injecting spoofed latitude and longitude at specific points during the trip to closely replicate actual GNSS spoofing incidents. The driving route used for testing is depicted in Figure 6. For turn-by-turn attacks, three instances are conducted along the route, involving shifts to new locations (adjacent blocks) three times, each representing separate events. Additionally, six wrong turn attacks are conducted, where three cases involve spoofed routes indicating left turns while the AV is actually taking right turns, and the other three cases represent spoofed routes indicating right turns while the AV is making left turns in reality. The turn-by-turn attack detection model achieves a perfect detection accuracy, successfully identifying all three location shifts with 100% accuracy.

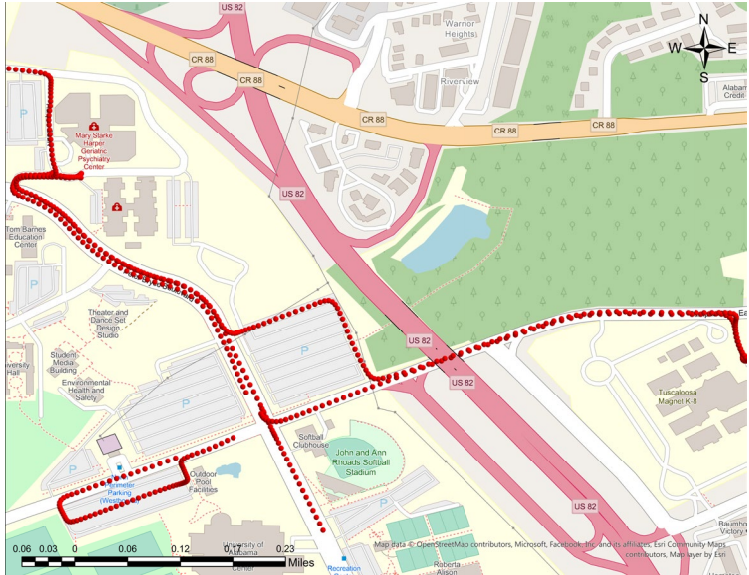

**Figure 6: Testing route**

**Table 5** presents the results of the wrong turn detection algorithm in terms of precision, recall, accuracy, and F1 score. Precision measures the accuracy of detecting wrong turn attacks among all the attack detection instances considered in this study. The precision for detecting wrong right turns is 95%, and for wrong left turns, it is 96%. Recall, on the other hand, indicates the percentage of observations where attacks are correctly detected among all the compromised observations. As shown in **Table 5**, the recall for wrong turns is 1, indicating that all wrong turn attacks are successfully identified. The accuracy of detecting wrong right turns is 95%, and for wrong left turns, it is 96%. The F1 score, which balances precision and recall, is 0.97 for wrong right turns and 0.98 for wrong left turns, demonstrating a well-balanced performance. Thus, the attack detection framework excels in real-time detection of wrong turn attacks, whether they involve right turns or left turns.

**TABLE 5 Random Forest model validation result**

| Turn type | precision | recall | accuracy | f1-score |
|-----------|-----------|--------|----------|----------|
| Right | 0.95 | 1 | 0.95 | 0.97 |
| Left | 0.96 | 1 | 0.96 | 0.98 |

**CONCLUSION**

This paper presents an experimental verification of a sensor fusion-based GNSS spoofing attack detection framework tailored for AVs. We model two types of spoofing attack scenarios and detail the attack creation process. The detection framework combines RTK GNSS receiver and high-performing IMU data to enhance AV navigation accuracy and reliability. The framework employs two distinct strategies: a LSTM neural network for predicting AV distance between timestamps, comparing it with inertial sensor-based location shifts and monitoring motion states; and an RF-supervised machine learning model for detecting and classifying turns using the steering angle sensor's output. GNSS-derived speed is compared with the actual vehicle velocity output in both strategies. A testing dataset is created to train the LSTM model for distance prediction and the RF model for turn classification. Validation results demonstrate both models' high accuracy in their respective tasks. Experiments conducted in Tuscaloosa, AL, simulating urban road structures, confirm the framework's effectiveness in detecting various GNSS spoofing attacks.

The results highlight the robustness and efficacy of the sensor fusion-based approach in safeguarding AVs against GNSS spoofing threats. By harnessing the integration of GNSS and IMU data, our proposed detection framework shows great potential in enhancing AV navigation security and resilience in real-time scenarios, thus advancing the safety and reliability of autonomous mobility.

## ACKNOWLEDGMENTS

This material is based on a study supported by National Science Foundation (NSF)— Award# **2244371 and Award # 2104999**. Any opinions, findings, and conclusions or recommendations expressed in this material are those of the author(s) and do not necessarily reflect the views of the NSF, and the NSF assumes no liability for the contents or use thereof.

## AUTHOR CONTRIBUTIONS

The authors confirm their contribution to the paper as follows: study conception and design: S. Dasgupta, K. Shakib, M. Rahman; data collection: S. Dasgupta, and K. Shakib; interpretation of results: S. Dasgupta, K. Shakib, and M. Rahman; draft manuscript preparation: S. Dasgupta, K. Shakib, and M. Rahman. All authors reviewed the results and approved the final version of the manuscript.

# REFERENCES

1. Dasgupta, S., M. Rahman, M. Islam, and M. Chowdhury. A Sensor Fusion-Based GNSS Spoofing Attack Detection Framework for Autonomous Vehicles. IEEE Transactions on Intelligent Transportation Systems, Vol. 23, No. 12, 2022, pp. 23559–23572. https://doi.org/10.1109/TITS.2022.3197817.

2. GPS.Gov. https://www.gps.gov/systems/gps/. Accessed Jul. 21, 2023.

3. BeiDou Navigation Satellite System. http://en.beidou.gov.cn/. Accessed Jul. 21, 2023.

4. GALILEO | European Global Navigation Satellite System. https://galileognss.eu/. Accessed Jul. 21, 2023.

5. About GLONASS. https://glonass-iac.ru/en/about_glonass/. Accessed Jul. 21, 2023.

6. IRNSS Programme. https://www.isro.gov.in/IRNSS_Programme.html. Accessed Jul. 21, 2023.

7. QZSS (Quasi-Zenith Satellite System) - Cabinet Office (Japan). https://qzss.go.jp/en/. Accessed Jul. 21, 2023.

8. SAPOS ® Precise Positioning in Location and Height Satellite Positioning Service of the Offical German Surveying and Mapping Agency for Geoinformation and State Survey of Lower Saxony (LGLN).

9. Zidan, J., E. I. Adegoke, E. Kampert, S. A. Birrell, C. R. Ford, and M. D. Higgins. GNSS Vulnerabilities and Existing Solutions: A Review of the Literature. IEEE Access, 2020, pp. 1–1. https://doi.org/10.1109/access.2020.2973759.

10. Adjrad, M., and P. Groves. 3D-Mapping-Aided GNSS Exploiting Galileo for Better Accuracy in Dense Urban Environments. 2017.

11. Adjrad, M., and P. D. Groves. Intelligent Urban Positioning Using Shadow Matching and GNSS Ranging Aided by 3D Mapping. No. 1, 2016, pp. 534–553.

12. Interference Effects And Mitigation Techniques. In Global Positioning System: Theory and Applications, Volume I, American Institute of Aeronautics and Astronautics, pp. 717–771.

13. Multipath Effects. In Global Positioning System: Theory and Applications, Volume I, American Institute of Aeronautics and Astronautics, pp. 547–568.

14. Zidan, J., E. I. Adegoke, E. Kampert, S. A. Birrell, C. R. Ford, and M. D. Higgins. GNSS Vulnerabilities and Existing Solutions: A Review of the Literature. IEEE Access, Vol. 9, 2021, pp. 153960–153976. https://doi.org/10.1109/ACCESS.2020.2973759.

15. Rügamer, A., F. Iis, and D. Kowalewski. Jamming and Spoofing of GNSS Signals-An Underestimated Risk?

16. Spoofing: Is Your GPS Attack Proof? | Septentrio. https://www.septentrio.com/en/learn-more/insights/spoofing-your-gps-attack-proof. Accessed Apr. 25, 2022.

17. Junzhi, L., L. Wanqing, F. Q. I. Research Progress of GNSS Spoofing and Spoofing Detection Technology. IEEE 19th International Conference on Communication, 2019.

18. Psiaki, M. L., and T. E. Humphreys. GNSS Spoofing and Detection. Proceedings of the IEEE, Vol. 104, No. 6, 2016, pp. 1258–1270. https://doi.org/10.1109/JPROC.2016.2526658.

19. Wu, Z., Y. Zhang, Y. Yang, C. Liang, and R. Liu. Spoofing and Anti-Spoofing Technologies of Global Navigation Satellite System: A Survey. IEEE Access, Vol. 8, 2020, pp. 165444–165496. https://doi.org/10.1109/access.2020.3022294.

20. Van Der Merwe, J. R., X. Zubizarreta, I. Lukčin, A. Rügamer, and W. Felber. Classification of Spoofing Attack Types. 2018 European Navigation Conference (ENC), 2018, pp. 91–99. https://doi.org/10.1109/EURONAV.2018.8433227.

21. Dasgupta, S., M. Rahman, M. Islam, and M. Chowdhury. Prediction-Based GNSS Spoofing Attack Detection for Autonomous Vehicles. Transportation Research Board, 2020.

22. Humphreys, T. E., B. M. Ledvina, M. L. Psiaki, B. W. O'Hanlon, P. M. Kintner, and Jr. Assessing the Spoofing Threat: Development of a Portable GPS Civilian Spoofer. 2008.

23. Warner, J. S., and R. G. Johnston. A Simple Demonstration That the Global Positioning System (GPS) Is Vulnerable to Spoofing. Journal of security administration, Vol. 25, No. 2, 2002, pp. 19–27.

24. Some, E., and A. J. Gasiewski. Software Defined Radio Injection-Locking Using a GPS Signal for Multichannel Coherent Receivers. IEEE Aerospace Conference Proceedings, Vol. 2023-March, 2023. https://doi.org/10.1109/AERO55745.2023.10115547.

25. Neish, A., S. Lo, Y. H. Chen, and P. Enge. Uncoupled Accelerometer Based GNSS Spoof Detection for Automobiles Using Statistic and Wavelet Based Tests. 2018.

26. D'Antonio, S., L. Coppolino, I. A. Elia, and V. Formicola. Security Issues of a Phasor Data Concentrator for Smart Grid Infrastructure. ACM International Conference Proceeding Series, 2011, pp. 3–8. https://doi.org/10.1145/1978582.1978584.

27. Almutairy, F., L. Scekic, M. Matar, R. Elmoudi, and S. Wshah. Detection and Mitigation of GPS Spoofing Attacks on Phasor Measurement Units Using Deep Learning. International Journal of Electrical Power & Energy Systems, Vol. 151, 2023, p. 109160. https://doi.org/10.1016/J.IJEPES.2023.109160.

28. MongoDB vs SQL Server: Which Is Better? [10 Critical Differences] - Learn | Hevo. https://hevodata.com/learn/mongodb-vs-sql-server/. Accessed Jul. 31, 2023.

29. Robusto, C. C. The Cosine-Haversine Formula. The American Mathematical Monthly, Vol. 64, No. 1, 1957, p. 38. https://doi.org/10.2307/2309088.

30. Khan, Z., M. Chowdhury, M. Islam, C. Y. Huang, and M. Rahman. Long Short-Term Memory Neural Network-Based Attack Detection Model for In-Vehicle Network Security. IEEE Sensors Letters, Vol. 4, No. 6, 2020. https://doi.org/10.1109/LSENS.2020.2993522.

31. Ramanishka, V., Y.-T. Chen, T. Misu, and K. Saenko. Toward Driving Scene Understanding: A Dataset for Learning Driver Behavior and Causal Reasoning. Proceedings of the IEEE Computer Society Conference on Computer Vision and Pattern Recognition, 2018, pp. 7699–7707.